\documentclass[11pt]{article}

\usepackage{amsfonts,amssymb,graphics,epsfig,verbatim,bm,latexsym,amsmath,url,amsbsy,
geometry}
\newtheorem{theorem}{Theorem}

\newtheorem{lemma}{Lemma}
\usepackage{csquotes}
\usepackage[authoryear]{natbib}
\usepackage{setspace}
\usepackage{float}
\restylefloat{table}
\bibpunct{(}{)}{;}{a}{,}{,}

\begin{document}
\bibliographystyle{plainnat}

\title{\bf A Comparison of First-Difference and Forward Orthogonal Deviations GMM}
\date{July 2019}
\author{Robert F. Phillips\footnote{Robert F. Phillips; address: 2115 G Street, NW, Suite 340, Washington DC, 20052; phone: 202-994-8619; fax: 202-994-6147; email: \texttt{rphil@gwu.edu}} \\George Washington University}

\maketitle

\begin{abstract}

This paper  provides a necessary and sufficient instruments condition  assuring two-step generalized method of moments (GMM) based on the forward orthogonal deviations transformation  is numerically equivalent to two-step GMM based on the first-difference transformation.  The  condition also tells us when system GMM, based on differencing, can be computed using forward orthogonal deviations.  Additionally, it tells us when forward orthogonal deviations and differencing do not lead to the same GMM estimator.  When estimators based on these two transformations differ, Monte Carlo simulations indicate that estimators based on forward orthogonal deviations have better finite sample properties than estimators based on differencing.

\end{abstract}


\section{Introduction \label{intro}}

A popular method for removing time-invariant effects from panel data is to first difference the data.  \cite{Arellano1991}, for example, proposed one-step and two-step first-difference generalized method of moments --- henceforth FD-GMM --- estimators. Later \cite{Bover1995} and \cite{Blundell1998} showed how equations in differences and equations in levels could be estimated as a system with generalized method of moments --- system GMM. Since that time, FD-GMM and system GMM have become dominant estimation approaches in the literature on panel data estimation.

The dominance of the first-difference transformation may, in part, be attributed to invariance results in \cite{Schmidt1992} and \cite{Bover1995}.  These results say that, under suitable restrictions, two different transformations can lead to the same generalized method of moments (GMM) estimator.  But if two different transformations lead to the same estimator, why bother with another transformation that gives the same result as the first-difference transformation?  Hence,  differencing may be all we need. 

It turns out, however, significant computational advantages may be possible using another transformation (see, e.g., Arellano and Bover, 1995; Phillips, 2019b). Moreover, it is not always the case that different transformations lead to the same GMM estimator. In particular, invariance to transformation depends on the instruments.  \cite{Schmidt1992} and \cite{Bover1995} focused on efficient estimation, and, in doing so, identified using all available instruments as sufficient for  invariance to transformation conclusions. But using all available instruments is not necessary for an invariance to transformation result \citep{Phillips2019a}.  \cite{Phillips2019a}, on the other hand, provided a sufficient {\em and necessary} instruments condition that assures a GMM estimator can be calculated using two-stage least squares (2SLS) after filtering the data. 

But the result in \cite{Phillips2019a} does not cover GMM when optimal weighting is used in the presence of conditional heteroskedasticity. This paper examines that case.  It shows that the  condition on the instruments identified in \cite{Phillips2019a} is necessary and sufficient for two-step FD-GMM to be equivalent to two-step GMM based on the forward orthogonal deviations transformation --- henceforth, two-step FOD-GMM. In fact, the result provided in this paper applies more generally than to the first difference and forward orthogonal deviations transformations. All we need assume about the transformation is that the transformation matrix $ \boldsymbol{K} $ that sweeps out the time-invariant effects is such that  $\boldsymbol{K}\boldsymbol{K}^{\prime}$ is a positive definite matrix.  Moreover, I show that if the instruments condition is met, and only if the instruments condition is met, then the well-known system GMM estimator (Arellano and Bover, 1995; Blundell and Bond, 1998) can be calculated using the forward orthogonal deviations transformation rather than the first-difference transformation.  

The necessity of the instruments condition tells us that, if a choice for instruments does not satisfy the condition, two different transformations of the data cannot lead to the same GMM estimator.  For example, experience has taught researchers that first-differencing and forward orthogonal deviations do not lead to the same GMM estimator when only recent lags are used as instrumental variables. The reason this is true is because the instruments condition is not satisfied \citep{Phillips2019a}.

But when different transformations must lead to different GMM estimators, as when only recent lags are used as instruments, the relevant question then becomes which transformation leads to the better GMM estimator?  This question has received some attention in the literature; see \cite{Hay2009}, \cite{Hsiao2017}, and \cite{Phillips2019b}. \cite{Hsiao2017} compared the asymptotic properties of method of moments estimators based on differencing the data versus using forward orthogonal deviations. That paper also provides some Monte Carlo evidence on the finite sample behavior of method of moments estimators based on differencing and on forward orthogonal deviations. \cite{Hay2009} examined the finite sample behavior of one-step GMM based on the forward orthogonal deviations transformation --- one-step FOD-GMM --- and one-step FD-GMM. He found that one-step FOD-GMM compared favorably to one-step FD-GMM.  And \cite{Phillips2019b} found that one-step FOD-GMM also outperformed two-step FD-GMM when the length of the time-series ($T$) is not small. 

In this paper I compare the finite sample properties of two-step FOD-GMM to two-step FD-GMM with Monte Carlo experiments.  I also investigate the finite sample properties of a system GMM estimator that exploits the forward orthogonal deviations transformation and compare its sampling behavior to that of the usual system GMM estimator, which relies on differencing.  I find that the estimators based on forward orthogonal deviations  dominate their counterparts based on differencing.  They generally have smaller absolute bias and their standard deviations are almost always smaller.

The next section provides numerical equivalence results for GMM based on different transformations. Section \ref{MC} provides the Monte Carlo evidence, and Section \ref{summary} concludes. Proofs are relegated to Section \ref{proofs}.

\section{Numerically equivalent transformations}

When panel data are used, the data are often transformed in order to remove time-invariant effects. Specifically, consider the model
\begin{equation}
\boldsymbol{y}_{i}=\boldsymbol{X}_{i}\boldsymbol{\beta }+\boldsymbol{\iota}\eta_{i}+\boldsymbol{v}_i,\text{ \ \ \ \ \ \ \ \ \ \ } i=1,\ldots
,N, \label{model}
\end{equation}%
where $\boldsymbol{X}_{i}$ is a matrix of observations on
explanatory variables, $\boldsymbol{v}_{i}$ is a  vector of errors that vary with time and individual, $\boldsymbol{\iota}$ is a vector of ones, and $\eta_i$ is an unobserved time-invariant effect.  The time-invariant effect $\eta_i$ can be removed by premultiplying through (\ref{model}) by a transformation matrix $\boldsymbol{K}$ that satisfies $\boldsymbol{K}\boldsymbol{\iota}=\boldsymbol{0}$.

Moreover, if $\boldsymbol{K}\boldsymbol{K}^{\prime}$ is a positive definite matrix, there exists another transformation that yields exactly the same estimator if, and only if, any instrument used in period $s$ can be constructed from a linear combination of instruments used for period $t$, for every $t \geq s$. This condition is satisfied in the well-known case where the instruments consist of lagged predetermined variables and all available instruments are used (see, e.g., Arellano, 2003, p. 153; Phillips, 2019a).  But other instrument choices also satisfy the instruments condition. The condition is satisfied whenever an instrument used in an earlier period is included, somehow, in the list of instruments used in a later period.  That possibility allows for many different choices of instruments that satisfy the instruments condition; see \cite{Phillips2019a} for examples.

The numerically equivalent transformation is provided in Theorem 1.
 
\begin{theorem} \label{thm1}
Let $\boldsymbol{z}_{it}$ be a $k_{t} \times 1$
vector of instruments $(t=1,\ldots ,R)$. Also, let%
\begin{equation}
\boldsymbol{Z}_{i}=\left( 
\begin{array}{cccc}
\boldsymbol{z}_{i1}^{\prime } & \boldsymbol{0} & \cdots & \boldsymbol{0} \\ 
\boldsymbol{0} & \boldsymbol{z}_{i2}^{\prime } & \cdots & \boldsymbol{0} \\ 
\vdots & \vdots & \ddots & \vdots \\ 
\boldsymbol{0} & \boldsymbol{0} & \cdots & \boldsymbol{z}_{iR}^{\prime }%
\end{array}%
\right) . \label{Z_i}
\end{equation}%
Moreover, let $\boldsymbol{K}$ be such that $\boldsymbol{K}\boldsymbol{\iota}=\boldsymbol{0}$ and $\boldsymbol{K}\boldsymbol{K}^{\prime}$ is positive definite. Let $\widehat{\boldsymbol{\beta}}$ be an initial estimator of $\boldsymbol{\beta}$ and set $\boldsymbol{e}_i=\boldsymbol{y}_i-\boldsymbol{X}_i\widehat{\boldsymbol{\beta}}$ $(i=1,\ldots,N)$. Furthermore, set $\boldsymbol{F}=\boldsymbol{U}\boldsymbol{K}$, where $\boldsymbol{U}$ is the upper-triangular Cholesky factorization of $(\boldsymbol{K}\boldsymbol{K}^{\prime})^{-1}$. Next, let $\boldsymbol{\tilde{y}}_i=\boldsymbol{K}\boldsymbol{y}_i$, $\boldsymbol{\tilde{X}}_i=\boldsymbol{K}\boldsymbol{X}_i$, and $\boldsymbol{\tilde{e}}_i=\boldsymbol{K}\boldsymbol{e}_i$ ($i=1,\ldots,N$). Also, set $\boldsymbol{\ddot{y}}_i=\boldsymbol{F}\boldsymbol{y}_i$, $\boldsymbol{\ddot{X}}_i=\boldsymbol{F}\boldsymbol{X}_i$, and $\boldsymbol{\ddot{e}}_i=\boldsymbol{F}\boldsymbol{e}_i$ ($i=1,\ldots,N$). Finally, define
\begin{eqnarray}
\widehat{\boldsymbol{\beta}}_K&=&\left[\sum_i\boldsymbol{\tilde{X}}_i^{\prime}\boldsymbol{Z}_i\left(\sum_i\boldsymbol{Z}_i^{\prime}\boldsymbol{\tilde{e}}_i\boldsymbol{\tilde{e}}_i^{\prime}\boldsymbol{Z}_i\right)^{-1}\sum_i\boldsymbol{Z}_i^{\prime}\boldsymbol{\tilde{X}}_i\right]^{-1} \nonumber \\
& & \times \sum_i\boldsymbol{\tilde{X}}_i^{\prime}\boldsymbol{Z}_i\left(\sum_i\boldsymbol{Z}_i^{\prime}\boldsymbol{\tilde{e}}_i\boldsymbol{\tilde{e}}_i^{\prime}\boldsymbol{Z}_i\right)^{-1}\sum_i\boldsymbol{Z}_i^{\prime}\boldsymbol{\tilde{y}}_i \label{beta_K}
\end{eqnarray}
and
\begin{eqnarray}
\widehat{\boldsymbol{\beta}}_F&=&\left[\sum_i\boldsymbol{\ddot{X}}_i^{\prime}\boldsymbol{Z}_i\left(\sum_i\boldsymbol{Z}_i^{\prime}\boldsymbol{\ddot{e}}_i\boldsymbol{\ddot{e}}_i^{\prime}\boldsymbol{Z}_i\right)^{-1}\sum_i\boldsymbol{Z}_i^{\prime}\boldsymbol{\ddot{X}}_i\right]^{-1} \nonumber \\
& & \times \sum_i\boldsymbol{\ddot{X}}_i^{\prime}\boldsymbol{Z}_i\left(\sum_i\boldsymbol{Z}_i^{\prime}\boldsymbol{\ddot{e}}_i\boldsymbol{\ddot{e}}_i^{\prime}\boldsymbol{Z}_i\right)^{-1}\sum_i\boldsymbol{Z}_i^{\prime}\boldsymbol{\ddot{y}}_i. \label{beta_F}
\end{eqnarray}
Then $\widehat{\boldsymbol{\beta}}_F=\widehat{\boldsymbol{\beta}}_K$  if, and only if, every entry in $\boldsymbol{z}_{is}$ is a linear combination of entries in $\boldsymbol{z}_{it}$ $(s = 1,\ldots, t$, $%
t=1,\ldots ,R)$.
\end{theorem}

See Section \ref{proofs} for a proof.

An important special case of Theorem \ref{thm1} is a first-differenced panel data model. In this case, $\boldsymbol{K}=\boldsymbol{D}$, where
\begin{equation*}
\boldsymbol{D}=\left( 
\begin{array}{ccccc}
-1 & 1 & 0 & \cdots  & 0 \\ 
0 & -1 & 1 & \cdots  & 0 \\ 
\vdots  & \vdots  & \ddots  & \ddots  & \vdots  \\ 
0 & 0 & \cdots  & -1 & 1%
\end{array}%
\right). \label{differencing}
\end{equation*}%
Given $\boldsymbol{K}=\boldsymbol{D}$, the appropriate $\boldsymbol{F}$ is the forward orthogonal deviations transformation matrix given by
\begin{eqnarray}
\boldsymbol{F} &=&\text{ diag}\left( \left( \frac{T-1}{T}\right)
^{1/2}, \left( \frac{T-2}{T-1}\right)
^{1/2}, \ldots ,\left( \frac{1}{2}\right) ^{1/2}\right)   \notag \\
&& \times \left( 
\begin{array}{ccccccc}
1 & - \frac{1}{T-1} & - \frac{1}{T-1} & \cdots & - \frac{1}{T-1} & - \frac{1}{T-1} & - \frac{1}{T-1} \\ 
0 & 1 & - \frac{1}{T-2} & \cdots & - \frac{1}{T-2} & - \frac{1}{T-2} & - \frac{1}{T-2} \\ 
\vdots & \vdots & \vdots &  & \vdots & \vdots & \vdots \\ 
0 & 0 & 0 & \cdots  & 1 & -\frac{1}{2} & -\frac{1}{2} \\ 
0 & 0 & 0 & \cdots & 0 & 1 & -1 \label{fod_trans}
\end{array}%
\right)  
\end{eqnarray}%
(see Arellano, 2003, p. 17). It is well-known that if the errors in $ \boldsymbol{v}_i $ are conditionally homoskedastic and uncorrelated, then the forward orthogonal deviation errors in $ \boldsymbol{\ddot{v}}_i = \boldsymbol{F}\boldsymbol{v}_i $ are conditionally homoskedastic and uncorrelated. However, even if the entries in $ \boldsymbol{v}_i $ are not conditionally homoskedastic and uncorrelated, according to Theorem \ref{thm1},  FD-GMM, based on its optimal weighting matrix, is equivalent to FOD-GMM, based on its optimal weighting matrix, if, and only if, every instrument used in period $s$ can be constructed as a linear combination of instruments used in period $t$, for $t \geq s$. 

In addition to being able to compute FD-GMM estimates in an alternative manner, the system GMM estimator has an alternative representation, provided the instruments condition is met.  As is well-known, the usual system GMM estimator uses both differenced and levels data.  However, system GMM estimates can alternatively be calculated using levels data and forward orthogonal deviations if, and only if, the instruments condition in Theorem \ref{thm1} is satisfied. 
 
To establish this claim, consider the model
 \begin{equation*}
y_{it}=\delta y_{i,t-1} + \boldsymbol{x}_{it}^{\prime}\boldsymbol{\alpha}+\eta_i + v_{it}, \text{ \ \ \ \ }t=1,\ldots,T, \text{ \ \ }i=1,\ldots, N.
\end{equation*}
Under suitable conditions, $\boldsymbol{\beta}=(\delta,\boldsymbol{\alpha}^{\prime})^{\prime}$ can be estimated with the system GMM estimator studied by \cite{Bover1995} and \cite{Blundell1998}.

In order to write that estimator, let $\boldsymbol{y}_i = (y_{i1},\ldots,y_{iT})^{\prime}$, and let $\boldsymbol{X}_i$ denote a  $T \times K$ matrix with $(y_{i,t-1},\boldsymbol{x}_{it}^{\prime})$ in its $t$th row ($t=1,\ldots, T$). Next set $\boldsymbol{y}_i^+ = (\boldsymbol{y}_i^{\prime},\boldsymbol{y}_i^{\prime})^{\prime}$ and $\boldsymbol{X}_i^+ = (\boldsymbol{X}_i^{\prime},\,\boldsymbol{X}_i^{\prime})^{\prime} $. The usual system GMM estimator relies on differencing the observations in the first $T$ rows in $\boldsymbol{y}_i^+$ and $\boldsymbol{X}_i^+$. Specifically, it uses the transformed data $\boldsymbol{\tilde{y}}_i^+=\boldsymbol{K}^+\boldsymbol{y}_i^+$ and $\boldsymbol{\tilde{X}}_i^+=\boldsymbol{K}^+\boldsymbol{X}_i^+$ ($i=1,\ldots,N$), where 
\begin{equation}
\boldsymbol{K}^+=\left( 
\begin{array}{cc}
\boldsymbol{D} & \boldsymbol{0}    \\ 
\boldsymbol{0} & \boldsymbol{I}  \\ 
\end{array}%
\right). \label{K+}
\end{equation}

For the instrument matrix, let $\boldsymbol{Z}_{1i}$ and $\boldsymbol{Z}_{2i}$ be block-diagonal instrument matrices, where $\boldsymbol{Z}_{1i}$ has $1\times k_t$ instrument vector $\boldsymbol{z}_{it}^{\prime}$ in its $t$th diagonal block ($t=1,\ldots,T-1$) and $\boldsymbol{Z}_{2i}$ has $\boldsymbol{x}_{i1}^{\prime}-\boldsymbol{x}_{i0}^{\prime}$ in its first diagonal block and  $({y}_{i,t-1}-y_{i,t-2},\boldsymbol{x}_{it}^{\prime}-\boldsymbol{x}_{i,t-1}^{\prime})$ in  diagonal blocks $t=2,\ldots,T$. Next, set
\begin{equation*}
\boldsymbol{Z}_i^+=\left( 
\begin{array}{cc}
\boldsymbol{Z}_{1i} & \boldsymbol{0}    \\ 
\boldsymbol{0} & \boldsymbol{Z}_{2i}  \\ 
\end{array}%
\right), 
\end{equation*} 

Given this transformation matrix and the preceding notation, the system GMM estimator can be expressed as
\begin{eqnarray}
\widehat{\boldsymbol{\beta}}_{K^+}&=&\left[\sum_i\boldsymbol{\tilde{X}}_i^{+ \prime}\boldsymbol{Z}_i^+\left(\sum_i\boldsymbol{Z}_i^{+ \prime}\boldsymbol{\tilde{e}}_i^+\boldsymbol{\tilde{e}}_i^{+ \prime}\boldsymbol{Z}_i^+\right)^{-1}\sum_i\boldsymbol{Z}_i^{+ \prime}\boldsymbol{\tilde{X}}_i^+\right]^{-1} \nonumber \\
& & \times \sum_i\boldsymbol{\tilde{X}}_i^{+ \prime}\boldsymbol{Z}_i^+\left(\sum_i\boldsymbol{Z}_i^{^+ \prime}\boldsymbol{\tilde{e}}_i^+\boldsymbol{\tilde{e}}_i^{+ \prime}\boldsymbol{Z}_i^+\right)^{-1}\sum_i\boldsymbol{Z}_i^{+ \prime}\boldsymbol{\tilde{y}}_i^+, \label{beta_K+}
\end{eqnarray}
where $\boldsymbol{\tilde{e}}_i^+=\boldsymbol{\tilde{y}}_i^+ - \boldsymbol{\tilde{X}}_i^+\widehat{\boldsymbol{\beta}}$ ($i=1,\ldots,N$) and $\widehat{\boldsymbol{\beta}}$ is an initial estimator of $\boldsymbol{\beta}$.

Alternatively, if the instruments condition is satisfied, the same system GMM estimator can be constructed using forward orthogonal deviations rather than first differences. In this case, the transformation matrix is
\begin{equation}
\boldsymbol{F}^+=\left( 
\begin{array}{cc}
\boldsymbol{F} & \boldsymbol{0}    \\ 
\boldsymbol{0} & \boldsymbol{I}  \\ 
\end{array}%
\right), \label{F+}
\end{equation}
where $\boldsymbol{F}$ is the forward orthogonal deviations transformation matrix given by Eq. (\ref{fod_trans}). Now let $\boldsymbol{\ddot{y}}_i^+=\boldsymbol{F}^+\boldsymbol{y}_i^+$ and $\boldsymbol{\ddot{X}}_i^+=\boldsymbol{F}^+\boldsymbol{X}_i^+$ ($i=1,\ldots,N$). Then define
\begin{eqnarray}
\widehat{\boldsymbol{\beta}}_{F^+}&=&\left[\sum_i\boldsymbol{\ddot{X}}_i^{+ \prime}\boldsymbol{Z}_i^+\left(\sum_i\boldsymbol{Z}_i^{+ \prime}\boldsymbol{\ddot{e}}_i^+\boldsymbol{\ddot{e}}_i^{+ \prime}\boldsymbol{Z}_i^+\right)^{-1}\sum_i\boldsymbol{Z}_i^{+ \prime}\boldsymbol{\ddot{X}}_i^+\right]^{-1} \nonumber \\
& & \times \sum_i\boldsymbol{\ddot{X}}_i^{+ \prime}\boldsymbol{Z}_i^+\left(\sum_i\boldsymbol{Z}_i^{^+ \prime}\boldsymbol{\ddot{e}}_i^+\boldsymbol{\ddot{e}}_i^{+ \prime}\boldsymbol{Z}_i^+\right)^{-1}\sum_i\boldsymbol{Z}_i^{+ \prime}\boldsymbol{\ddot{y}}_i^+, \label{beta_F+}
\end{eqnarray}
where $\boldsymbol{\ddot{e}}_i^+=\boldsymbol{\ddot{y}}_i^+ - \boldsymbol{\ddot{X}}_i^+\widehat{\boldsymbol{\beta}}$ ($i=1,\ldots,N$).

We can now state Theorem \ref{thm2}.

\begin{theorem}\label{thm2}
Suppose $\widehat{\boldsymbol{\beta}}_{K^+}$ and $\widehat{\boldsymbol{\beta}}_{F^+}$ use the same initial estimator $\widehat{\boldsymbol{\beta}}$. Then $\widehat{\boldsymbol{\beta}}_{F^+}=\widehat{\boldsymbol{\beta}}_{K^+}$ if, and only if, every entry in $\boldsymbol{z}_{is}$ is a linear combination of entries in $\boldsymbol{z}_{it}$ $(s = 1,\ldots, t$, $%
t=1,\ldots ,T-1)$.
\end{theorem}

The proof is provided in Section \ref{proofs}.

Theorem \ref{thm1} applies to an important case. Specifically, it tells us when system GMM based on first differences is equivalent to system GMM based on forward orthogonal deviations. But the result holds more generally. In particular, we can replace $\boldsymbol{D}$ in the definition of $\boldsymbol{K}^+$ with another transformation matrix $\boldsymbol{K}$, provided $\boldsymbol{K}\boldsymbol{K}^{\prime}$ is positive definite and provided $\boldsymbol{F}$ in $\boldsymbol{F}^+$ is given by $\boldsymbol{F}=\boldsymbol{U}\boldsymbol{K}$, where $\boldsymbol{U}$ is the upper-triangular Cholesky factorization of $(\boldsymbol{K}\boldsymbol{K}^{\prime})^{-1}$.

\section{When only recent lags are used as instruments \label{MC}}

Theorem \ref{thm1} not only tells us when two different transformations lead to the same GMM estimator, it also tells us when they do not. For example, it tells us we cannot use two different transformations to get the same GMM estimator for a popular choice for instruments --- specifically, when only recent lags of predetermined variables are used as instruments. This is because, if only recent lags are used as instrumental variables, a lagged predetermined variable that is used as an instrument in an earlier period will not be used as an instrument in some later period, and consequently we cannot construct the instrument used in the earlier period from the instruments used in the later period.  In other words, the instruments condition is violated. Similarly, Theorem \ref{thm2} tells us that when only recent lags of predetermined variables are used as instruments, the system GMM estimator based on first differences --- the FD-SYS estimator --- is not the same as the system GMM estimator that exploits forward orthogonal deviations --- the FOD-SYS  estimator. 

But these observations raise some questions. When two-step FOD-GMM is not the same as two-step FD-GMM, yet both rely on the same choice of instruments and both are based on their respective optimal weighting matrices, which estimator is the better choice? Also, when FD-SYS and FOD-SYS estimators are not the same, which system estimator should we use? 

This section addresses these questions with Monte Carlo experiments.

\subsection{Monte Carlo simulations}

The experiments conducted for this paper are similar to those used in \cite{Phillips2019b}.  This allows the reader to compare the finite sample behavior of the estimators examined here to that of the estimators studied in \cite{Phillips2019b}.  

For all of the Monte Carlo simulations, data were generated according to the model
\begin{equation*}
y_{it}=\delta y_{i,t-1}+\alpha x_{it}+\eta _{i}+v_{it}%
\text{, \ \ \ \ \ \ \ \ \ }t=-49,\ldots ,T,\ i=1,\ldots ,N.
\end{equation*}%
Moreover,  the $x_{it}$s were generated as predetermined variables: 
\begin{equation*}
x_{it}=\rho x_{i,t-1} - 0.3 y_{i,t-1}+0.5\eta _{i}+\xi _{it},\text{ \ \ \ \ \
	\ \ \ \ \ \ \ \ }t=-49,\ldots ,T,\text{ \ }i=1,\ldots ,N.
\end{equation*}%

The start-up values $y_{i,-50}$ and $x_{i,-50}$ were set as  $y_{i,-50}=0$ and $x_{i,-50}=5+10\xi _{i,-50}$. Moreover, start-up
observations were discarded.  In particular, for each sample, estimation was based on the $T+1$
observations $\left( x_{i0},y_{i0}\right) ,\ldots ,\left(
x_{iT},y_{iT}\right) $ ($i=1,\ldots ,N$), with $N$ always set to 200 and $T$ set to either 10 or 30. For each sample size and combination of parameters, 10,000 independent samples were drawn.

In order to generate a sample, the parameters had to be specified and pseudo random numbers were generated.  For the parameters, I set $\alpha=0.5$, and $\delta$ was either 0.5 or 0.9, while $\rho$ was either 0.3 or 0.8. As for the pseudo random variates, the $\xi _{it}$s
were generated as independent uniform random variates with mean zero and
variance one. Moreover, the individual-specific effects --- the $\eta _{i}$s --- were generated independently of the $\xi _{it}$s and $v_{it}$s
as $\eta _{i}=\sigma _{\eta }\zeta _{i}$ ($i=1,\ldots,N$), with $%
\zeta _{i}$ a standard normal random variable. Two values for the
standard deviation $\sigma _{\eta }$ were considered: one or four.  

Moreover, two models were used to generate the $v_{it}$s: a conditionally heteroskedastic errors model and a time-series heteroskedastic errors model. For conditionally heteroskedastic errors, I set $v_{it}=x_{it} \epsilon
_{it}$, with $\epsilon _{it}$ a standard normal random variable, which was generated independently of $%
x_{it}$, $x_{i,t-s}$, and $\epsilon _{i,t-s}$ for $s\geq 1$. For time-series heteroskedastic errors,  $\lambda_{t}$s ($t=1, \ldots, T$) were first generated as uniform random variates with mean zero and variance one. Then I set $v_{it} = \lambda_{t} \epsilon_{it}$.  

\subsection{Results}

To get some sense of magnitudes, Table \ref{fd_bias_sd_rmse} provides bias, standard deviation, and root mean squared error estimates for the two-step FD-GMM  estimator.\footnote{All computations were performed using GAUSS.} The results in Table \ref{fd_bias_sd_rmse} are for an estimator that uses only recent lags of the predetermined explanatory variables as instruments. In particular, for $ \boldsymbol{z}_{it} $, I set $ \boldsymbol{z}_{i1}^{\prime} = (y_{i0},x_{i0},x_{i1}) $ and $ \boldsymbol{z}_{it}^{\prime} = (y_{i,t-2},y_{i,t-1}, x_{i,t-2},x_{i,t-1},x_{it}) $ ($ t=2,\ldots,T-1 $). To compute two-step FD-GMM estimates I used the formula in (\ref{beta_K}) with $\boldsymbol{K} = \boldsymbol{D}$.  These estimates require that one-step estimates first be calculated, and for those I used the formula in (\ref{beta_K}) with the weighting matrix $ \left(\sum_i\boldsymbol{Z}_i^{\prime}\boldsymbol{\tilde{e}}_i\boldsymbol{\tilde{e}}_i^{\prime}\boldsymbol{Z}_i\right)^{-1} $  replaced by $\left(\sum_i\boldsymbol{Z}_i^{\prime}\boldsymbol{D}\boldsymbol{D}^{\prime}\boldsymbol{Z}_i\right)^{-1}$.

Table \ref{fod_vs_fd} shows how the two-step FOD-GMM estimator compares to the two-step FD-GMM estimator when only recent lags of predetermined explanatory variables  are used as instruments.\footnote{The two-step FOD-GMM estimator is based on the same instruments as the FD-GMM estimator. The two-step FOD-GMM estimator is given by  (\ref{beta_F}) with $ \boldsymbol{F} $ given by the transformation matrix in (\ref{fod_trans}). Like two-step FD-GMM estimates, two-step FOD-GMM estimates require one-step estimates first be calculated. For the two-step FOD-GMM estimates, the one-step estimates were one-step FOD-GMM estimates.}  The table gives the percent reduction in absolute bias, standard deviation, and root mean squared error from using the FOD-GMM estimator rather than the FD-GMM estimator.  Specifically, the estimates in Table \ref{fod_vs_fd} were calculated as
\begin{equation*}
\frac{100(FD-FOD)}{FD},
\end{equation*}
where $ FD $ and $ FOD $ stand for the absolute bias, standard deviation, or root mean squared error of the two-step FD-GMM estimator and the two-step FOD-GMM estimator, respectively.

It is clear from the data in Table \ref{fod_vs_fd} that for the vast majority of sample designs the FOD-GMM estimator  has smaller absolute bias than the FD-GMM estimator.  There are only eight cases for which the FD-GMM estimator has the smaller absolute bias --- i.e., cases for which the percent reduction in bias is negative.  And, for most of those cases, the bias of both estimators is small relative to their standard deviations.  This is obvious from the fact that even though the FOD-GMM estimator has larger bias in these cases, for most of these cases its percentage reduction in root mean squared error is similar to its percentage reduction in standard deviation, which indicates that the standard deviations contribute more to the root mean squared errors than the biases of the two estimators. 

The data in Table \ref{fod_vs_fd} also reveal that the two-step FOD-GMM estimator is almost always the more efficient estimator. The percent reduction in standard deviation is almost always positive. There is only one case for which it is not positive. And, for that case, the FOD-GMM estimator loses little efficiency relative to the FD-GMM estimator. This fact and the fact that the FOD-GMM estimator usually has smaller bias than the FD-GMM estimator implies that, when precision is measured in terms of root mean squared error, the FOD-GMM estimator is always the more precise estimator.

Table \ref{fdsys_bias_sd_rmse} provides bias, standard deviation, and root mean squared error estimates for the FD-SYS estimator. This estimator is given by Eq. (\ref{beta_K+}) with $ \boldsymbol{K}^+ $ given by Eq. (\ref{K+}). The results in Table \ref{fdsys_bias_sd_rmse} are limited to $T = 10$ because for $T=30$ there are so many moment restrictions that  $ \sum_i\boldsymbol{Z}_i^{+ \prime}\boldsymbol{\ddot{e}}_i^+\boldsymbol{\ddot{e}}_i^{+ \prime}\boldsymbol{Z}_i^+ $ is singular, and, therefore, the optimal weighting matrix cannot be computed.

Table \ref{fodsys_vs_fdsys} reports the percent reduction in absolute bias, standard deviation, and root mean squared error provided by the FOD-SYS estimator. The FOD-SYS estimator is given by Eq. (\ref{beta_F+}) with  $ \boldsymbol{F}^+ $ given by Eq. (\ref{F+}).  The message of Table \ref{fodsys_vs_fdsys} is similar to that of Table \ref{fod_vs_fd}. The FOD-SYS estimator typically has smaller absolute bias, it is always more efficient, and it also always has the smaller root mean squared error.

	\begin{table}[H]
	\caption{Bias, standard deviation, and root mean squared error estimates for two-step FD-GMM  ($N=$ $200$).}
	\centering
	\begin{tabular}{lcccccccccc} 
		\hline
		&  & &  & &  \multicolumn{3}{c}{$\delta $ estimator} &    \multicolumn{3}{c}{$\alpha$ estimator}    \\ 
		&  $ T $ & $\sigma _{\eta }$ & $\delta$ & $\rho$ & bias & (sd) &
		\textit{rmse} &   bias & (sd) & \textit{rmse}  \\ \hline
		&  &  &  &  &  &  &   &  &  & \\ 
		
		\multicolumn{6}{l}{\underline{Conditionally heteroskedastic errors:}} & & & &  \\
		
		&  &  &  &  &  &  &   &  &   &    \\
		
		& $10$ &  1 & 0.5 & 0.3 &  \multicolumn{1}{r}{$-0.0173$} & \multicolumn{1}{r}{($0.0404$)} &
		\multicolumn{1}{r}{$\mathit{0.0440}$} &  
		\multicolumn{1}{r}{$-0.0029$} &  \multicolumn{1}{r}{($0.0433$)} & \multicolumn{1}{r}{$\mathit{0.0434}$}   \\ 
		
		& $10$ & 1 & 0.5 & 0.8 &  \multicolumn{1}{r}{$-0.0147$} & \multicolumn{1}{r}{($0.0333$)} &
		\multicolumn{1}{r}{$\mathit{0.0364}$} &  
		\multicolumn{1}{r}{$0.0325$} &  \multicolumn{1}{r}{($0.0526$)} & \multicolumn{1}{r}{$\mathit{0.0618}$}   \\ 
		
		& $10$ & 1 & 0.9 & 0.3 &  \multicolumn{1}{r}{$-0.0942$} & \multicolumn{1}{r}{($0.0929$)} &
		\multicolumn{1}{r}{$\mathit{0.1323}$} &  
		\multicolumn{1}{r}{$-0.0240$} &  \multicolumn{1}{r}{($0.0751$)} & \multicolumn{1}{r}{$\mathit{0.0789}$}   \\ 
		
		& $10$ & 1 & 0.9 & 0.8 &  \multicolumn{1}{r}{$-0.0499$} & \multicolumn{1}{r}{($0.1329$)} &
		\multicolumn{1}{r}{$\mathit{0.1420}$} &  
		\multicolumn{1}{r}{$0.1116$} &  \multicolumn{1}{r}{($0.2351$)} & \multicolumn{1}{r}{$\mathit{0.2602}$}  \\ 	
		
		& $10$ & 4 & 0.5 & 0.3 &  \multicolumn{1}{r}{$-0.0442$} & \multicolumn{1}{r}{($0.0631$)} &
		\multicolumn{1}{r}{$\mathit{0.0770}$} &  
		\multicolumn{1}{r}{$-0.0164$} &  \multicolumn{1}{r}{($0.0536$)} & \multicolumn{1}{r}{$\mathit{0.0560}$}   \\

		& $10$ & 4 & 0.5 & 0.8 &  \multicolumn{1}{r}{$-0.0254$} & \multicolumn{1}{r}{($0.0464$)} &
		\multicolumn{1}{r}{$\mathit{0.0529}$} &  
		\multicolumn{1}{r}{$0.0317$} &  \multicolumn{1}{r}{($0.0609$)} & \multicolumn{1}{r}{$\mathit{0.0687}$}   \\ 
		
		& $10$ & 4 & 0.9 & 0.3 &  \multicolumn{1}{r}{$-0.1622$} & \multicolumn{1}{r}{($0.1768$)} &
		\multicolumn{1}{r}{$\mathit{0.2400}$} &  
		\multicolumn{1}{r}{$-0.0867$} &  \multicolumn{1}{r}{($0.2566$)} & \multicolumn{1}{r}{$\mathit{0.2709}$}   \\

		& $10$ & 4 & 0.9 & 0.8 &  \multicolumn{1}{r}{$-0.0521$} & \multicolumn{1}{r}{($0.1512$)} &
		\multicolumn{1}{r}{$\mathit{0.1599}$} &  
		\multicolumn{1}{r}{$0.1470$} &  \multicolumn{1}{r}{($0.2634$)} & \multicolumn{1}{r}{$\mathit{0.3017}$}   \\

		&  &  &  &  &  &  &   &  &  & \\ 
		
		& $30$ &  1 & 0.5 & 0.3 &  \multicolumn{1}{r}{$-0.0085$} & \multicolumn{1}{r}{($0.0187$)} &
		\multicolumn{1}{r}{$\mathit{0.0206}$} &  
		\multicolumn{1}{r}{$-0.0004$} &  \multicolumn{1}{r}{($0.0231$)} & \multicolumn{1}{r}{$\mathit{0.0231}$}   \\ 
		
		& $30$ & 1 & 0.5 & 0.8 &  \multicolumn{1}{r}{$-0.0063$} & \multicolumn{1}{r}{($0.0192$)} &
		\multicolumn{1}{r}{$\mathit{0.0202}$} &  
		\multicolumn{1}{r}{$0.0184$} &  \multicolumn{1}{r}{($0.0278$)} & \multicolumn{1}{r}{$\mathit{0.0333}$}   \\ 
		
		& $30$ & 1 & 0.9 & 0.3 &  \multicolumn{1}{r}{$-0.0274$} & \multicolumn{1}{r}{($0.0368$)} &
		\multicolumn{1}{r}{$\mathit{0.0458}$} &  
		\multicolumn{1}{r}{$0.0032$} &  \multicolumn{1}{r}{($0.0377$)} & \multicolumn{1}{r}{$\mathit{0.0378}$}   \\ 
		
		& $30$ & 1 & 0.9 & 0.8 &  \multicolumn{1}{r}{$-0.0121$} & \multicolumn{1}{r}{($0.0751$)} &
		\multicolumn{1}{r}{$\mathit{0.0761}$} &  
		\multicolumn{1}{r}{$0.1102$} &  \multicolumn{1}{r}{($0.1468$)} & \multicolumn{1}{r}{$\mathit{0.1835}$}   \\

		& $30$ & 4 & 0.5 & 0.3 &  \multicolumn{1}{r}{$-0.0184$} & \multicolumn{1}{r}{($0.0252$)} &
		\multicolumn{1}{r}{$\mathit{0.0313}$} &  
		\multicolumn{1}{r}{$-0.0050$} &  \multicolumn{1}{r}{($0.0258$)} & \multicolumn{1}{r}{$\mathit{0.0263}$}   \\ 
		
		& $30$ & 4 & 0.5 & 0.8 &  \multicolumn{1}{r}{$-0.0098$} & \multicolumn{1}{r}{($0.0257$)} &
		\multicolumn{1}{r}{$\mathit{0.0275}$} &  
		\multicolumn{1}{r}{$0.0208$} &  \multicolumn{1}{r}{($0.0328$)} & \multicolumn{1}{r}{$\mathit{0.0388}$}   \\ 
		
		& $30$ & 4 & 0.9 & 0.3 &  \multicolumn{1}{r}{$-0.0364$} & \multicolumn{1}{r}{($0.0777$)} &
		\multicolumn{1}{r}{$\mathit{0.0858}$} &  
		\multicolumn{1}{r}{$0.0365$} &  \multicolumn{1}{r}{($0.1412$)} & \multicolumn{1}{r}{$\mathit{0.1459}$}   \\ 
		
		& $30$ & 4 & 0.9 & 0.8 &  \multicolumn{1}{r}{$-0.0102$} & \multicolumn{1}{r}{($0.0710$)} &
		\multicolumn{1}{r}{$\mathit{0.0718}$} &  
		\multicolumn{1}{r}{$0.1330$} &  \multicolumn{1}{r}{($0.1539$)} & \multicolumn{1}{r}{$\mathit{0.2034}$}   \\ 
		
		&  &  &  &  &  &  &   &  &  & \\ 
		
		\multicolumn{6}{l}{\underline{Time-series heteroskedastic errors:}} & & & &  \\
		
		&  &  &  &  &  &  &   &  &  & \\ 
		
		& $10$ &  1 & 0.5 & 0.3 &  \multicolumn{1}{r}{$-0.0062$} & \multicolumn{1}{r}{($0.0265$)} &
		\multicolumn{1}{r}{$\mathit{0.0272}$} &  
		\multicolumn{1}{r}{$-0.0016$} &  \multicolumn{1}{r}{($0.0234$)} & \multicolumn{1}{r}{$\mathit{0.0234}$}   \\ 
		
		& $10$ & 1 & 0.5 & 0.8 &  \multicolumn{1}{r}{$-0.0034$} & \multicolumn{1}{r}{($0.0195$)} &
		\multicolumn{1}{r}{$\mathit{0.0198}$} &  
		\multicolumn{1}{r}{$0.0010$} &  \multicolumn{1}{r}{($0.0207$)} & \multicolumn{1}{r}{$\mathit{0.0207}$}   \\ 
		
		& $10$ & 1 & 0.9 & 0.3 &  \multicolumn{1}{r}{$-0.0230$} & \multicolumn{1}{r}{($0.0423$)} &
		\multicolumn{1}{r}{$\mathit{0.0481}$} &  
		\multicolumn{1}{r}{$-0.0107$} &  \multicolumn{1}{r}{($0.0289$)} & \multicolumn{1}{r}{$\mathit{0.0308}$}   \\ 
		
		& $10$ & 1 & 0.9 & 0.8 &  \multicolumn{1}{r}{$-0.0026$} & \multicolumn{1}{r}{($0.0145$)} &
		\multicolumn{1}{r}{$\mathit{0.0148}$} &  
		\multicolumn{1}{r}{$0.0004$} &  \multicolumn{1}{r}{($0.0151$)} & \multicolumn{1}{r}{$\mathit{0.0151}$}   \\

		& $10$ & 4 & 0.5 & 0.3 &  \multicolumn{1}{r}{$-0.0143$} & \multicolumn{1}{r}{($0.0365$)} &
		\multicolumn{1}{r}{$\mathit{0.0392}$} &  
		\multicolumn{1}{r}{$-0.0057$} &  \multicolumn{1}{r}{($0.0275$)} & \multicolumn{1}{r}{$\mathit{0.0281}$}   \\

		& $10$ & 4 & 0.5 & 0.8 &  \multicolumn{1}{r}{$-0.0059$} & \multicolumn{1}{r}{($0.0246$)} &
		\multicolumn{1}{r}{$\mathit{0.0253}$} &  
		\multicolumn{1}{r}{$0.0004$} &  \multicolumn{1}{r}{($0.0225$)} & \multicolumn{1}{r}{$\mathit{0.0225}$}   \\ 
		
		& $10$ & 4 & 0.9 & 0.3 &  \multicolumn{1}{r}{$-0.0568$} & \multicolumn{1}{r}{($0.0698$)} &
		\multicolumn{1}{r}{$\mathit{0.0900}$} &  
		\multicolumn{1}{r}{$-0.0279$} &  \multicolumn{1}{r}{($0.0414$)} & \multicolumn{1}{r}{$\mathit{0.0500}$}   \\

		& $10$ & 4 & 0.9 & 0.8 &  \multicolumn{1}{r}{$-0.0065$} & \multicolumn{1}{r}{($0.0222$)} &
		\multicolumn{1}{r}{$\mathit{0.0232}$} &  
		\multicolumn{1}{r}{$-0.0018$} &  \multicolumn{1}{r}{($0.0172$)} & \multicolumn{1}{r}{$\mathit{0.0173}$}   \\

		&  &  &  &  &  &  &   &  &  & \\ 
		
		& $30$ &  1 & 0.5 & 0.3 &  \multicolumn{1}{r}{$-0.0050$} & \multicolumn{1}{r}{($0.0144$)} &
		\multicolumn{1}{r}{$\mathit{0.0153}$} &  
		\multicolumn{1}{r}{$-0.0006$} &  \multicolumn{1}{r}{($0.0139$)} & \multicolumn{1}{r}{$\mathit{0.0139}$}   \\ 
		
		& $30$ & 1 & 0.5 & 0.8 &  \multicolumn{1}{r}{$-0.0020$} & \multicolumn{1}{r}{($0.0112$)} &
		\multicolumn{1}{r}{$\mathit{0.0114}$} &  
		\multicolumn{1}{r}{$0.0018$} &  \multicolumn{1}{r}{($0.0123$)} & \multicolumn{1}{r}{$\mathit{0.0125}$}   \\ 
		
		& $30$ & 1 & 0.9 & 0.3 &  \multicolumn{1}{r}{$-0.0095$} & \multicolumn{1}{r}{($0.0169$)} &
		\multicolumn{1}{r}{$\mathit{0.0193}$} &  
		\multicolumn{1}{r}{$-0.0025$} &  \multicolumn{1}{r}{($0.0142$)} & \multicolumn{1}{r}{$\mathit{0.0144}$}   \\ 
		
		& $30$ & 1 & 0.9 & 0.8 &  \multicolumn{1}{r}{$-0.0011$} & \multicolumn{1}{r}{($0.0076$)} &
		\multicolumn{1}{r}{$\mathit{0.0077}$} &  
		\multicolumn{1}{r}{$0.0016$} &  \multicolumn{1}{r}{($0.0085$)} & \multicolumn{1}{r}{$\mathit{0.0086}$}   \\

		& $30$ & 4 & 0.5 & 0.3 &  \multicolumn{1}{r}{$-0.0104$} & \multicolumn{1}{r}{($0.0183$)} &
		\multicolumn{1}{r}{$\mathit{0.0210}$} &  
		\multicolumn{1}{r}{$-0.0034$} &  \multicolumn{1}{r}{($0.0149$)} & \multicolumn{1}{r}{$\mathit{0.0153}$}   \\ 
		
		& $30$ & 4 & 0.5 & 0.8 &  \multicolumn{1}{r}{$-0.0043$} & \multicolumn{1}{r}{($0.0132$)} &
		\multicolumn{1}{r}{$\mathit{0.0140}$} &  
		\multicolumn{1}{r}{$0.0007$} &  \multicolumn{1}{r}{($0.0127$)} & \multicolumn{1}{r}{$\mathit{0.0127}$}   \\ 
		
		& $30$ & 4 & 0.9 & 0.3 &  \multicolumn{1}{r}{$-0.0368$} & \multicolumn{1}{r}{($0.0309$)} &
		\multicolumn{1}{r}{$\mathit{0.0480}$} &  
		\multicolumn{1}{r}{$-0.0164$} &  \multicolumn{1}{r}{($0.0196$)} & \multicolumn{1}{r}{$\mathit{0.0256}$}   \\ 
		
		& $30$ & 4 & 0.9 & 0.8 &  \multicolumn{1}{r}{$-0.0038$} & \multicolumn{1}{r}{($0.0104$)} &
		\multicolumn{1}{r}{$\mathit{0.0110}$} &  
		\multicolumn{1}{r}{$0.0002$} &  \multicolumn{1}{r}{($0.0093$)} & \multicolumn{1}{r}{$\mathit{0.0093}$}   \\

		&  &  &  &  &  &  &   &   &  &    \\ \hline
		
	\end{tabular}	 
	\label{fd_bias_sd_rmse}
\end{table}

\begin{table}[H]
	\caption{Percent reduction in absolute bias, standard deviation, and root mean squared error estimates for two-step FOD-GMM relative to two-step FD-GMM (i.e., 100(FD $-$ FOD)/FD).}
	\centering
	\begin{tabular}{lcccccccccccc}
		\hline
		& & & &  & &  \multicolumn{3}{c}{$\delta $ estimator} & &   \multicolumn{3}{c}{$\alpha$ estimator}    \\ 
		& $T$ &  $\sigma _{\eta }$ & $\delta$ & $\rho$ & & bias & (sd) &
		\textit{rmse} & &  bias & (sd) & \textit{rmse}  \\ \hline
		& & &  &  &  &  &  &   &  & & &  \\
		
		\multicolumn{8}{l}{\underline{Conditionally heteroskedastic errors:}} & & & &   \\
		
		& & &  &  &  &  &  &   &  & & &  \\
		
		& 10 &  1 & 0.5 & 0.3 & & \multicolumn{1}{r}{$26.7$} & \multicolumn{1}{r}{($4.9$)} &
		\multicolumn{1}{r}{$\mathit{7.9}$} & & 
		\multicolumn{1}{r}{$92.8$} &  \multicolumn{1}{r}{($5.5$)} & \multicolumn{1}{r}{$\mathit{5.7}$}   \\

		& 10 &  1 & 0.5 & 0.8 & & \multicolumn{1}{r}{$24.0$} & \multicolumn{1}{r}{($10.8$)} &
		\multicolumn{1}{r}{$\mathit{12.8}$} & & 
		\multicolumn{1}{r}{$3.6$} &  \multicolumn{1}{r}{($4.6$)} & \multicolumn{1}{r}{$\mathit{4.4}$}   \\

		& 10 &  1 & 0.9 & 0.3 & & \multicolumn{1}{r}{$29.5$} & \multicolumn{1}{r}{($17.9$)} &
		\multicolumn{1}{r}{$\mathit{23.5}$} &  &
		\multicolumn{1}{r}{$71.2$} &  \multicolumn{1}{r}{($14.5$)} & \multicolumn{1}{r}{$\mathit{18.1}$}   \\

		& 10 &  1 & 0.9 & 0.8 & & \multicolumn{1}{r}{$9.6$} & \multicolumn{1}{r}{($11.1$)} &
		\multicolumn{1}{r}{$\mathit{10.9}$} &  & 
		\multicolumn{1}{r}{$15.7$} &  \multicolumn{1}{r}{($7.2$)} & \multicolumn{1}{r}{$\mathit{8.7}$}   \\

		& 10 &  4 & 0.5 & 0.3 & & \multicolumn{1}{r}{$70.3$} & \multicolumn{1}{r}{($27.9$)} &
		\multicolumn{1}{r}{$\mathit{38.5}$} & & 
		\multicolumn{1}{r}{$89.1$} &  \multicolumn{1}{r}{($16.9$)} & \multicolumn{1}{r}{$\mathit{20.5}$}   \\

		& 10 &  4 & 0.5 & 0.8 & & \multicolumn{1}{r}{$53.6$} & \multicolumn{1}{r}{($19.0$)} &
		\multicolumn{1}{r}{$\mathit{25.5}$} &  &
		\multicolumn{1}{r}{$-19.2$} &  \multicolumn{1}{r}{($7.5$)} & \multicolumn{1}{r}{$\mathit{1.3}$}   \\

		& 10 &  4 & 0.9 & 0.3 & & \multicolumn{1}{r}{$51.0$} & \multicolumn{1}{r}{($21.5$)} &
		\multicolumn{1}{r}{$\mathit{33.4}$} &  &
		\multicolumn{1}{r}{$98.4$} &  \multicolumn{1}{r}{($10.3$)} & \multicolumn{1}{r}{$\mathit{15.1}$}   \\

		& 10 &  4 & 0.9 & 0.8 & & \multicolumn{1}{r}{$13.1$} & \multicolumn{1}{r}{($13.5$)} &
		\multicolumn{1}{r}{$\mathit{13.5}$} &  &
		\multicolumn{1}{r}{$18.0$} &  \multicolumn{1}{r}{($9.0$)} & \multicolumn{1}{r}{$\mathit{11.0}$}   \\

		& &  &  &  &  &  &  &   &  & & &  \\
		
		& 30 &  1 & 0.5 & 0.3 & & \multicolumn{1}{r}{$52.7$} & \multicolumn{1}{r}{($1.7$)} &
		\multicolumn{1}{r}{$\mathit{8.4}$} & & 
		\multicolumn{1}{r}{$-11.5$} &  \multicolumn{1}{r}{($5.7$)} & \multicolumn{1}{r}{$\mathit{5.7}$}   \\

		& 30 &  1 & 0.5 & 0.8 & & \multicolumn{1}{r}{$32.0$} & \multicolumn{1}{r}{($13.6$)} &
		\multicolumn{1}{r}{$\mathit{15.2}$} & & 
		\multicolumn{1}{r}{$25.1$} &  \multicolumn{1}{r}{($5.2$)} & \multicolumn{1}{r}{$\mathit{10.8}$}   \\

		& 30 &  1 & 0.9 & 0.3 & & \multicolumn{1}{r}{$27.5$} & \multicolumn{1}{r}{($5.6$)} &
		\multicolumn{1}{r}{$\mathit{12.8}$} &  &
		\multicolumn{1}{r}{$-14.1$} &  \multicolumn{1}{r}{($17.0$)} & \multicolumn{1}{r}{$\mathit{16.7}$}   \\

		& 30 &  1 & 0.9 & 0.8 & & \multicolumn{1}{r}{$-46.7$} & \multicolumn{1}{r}{($8.3$)} &
		\multicolumn{1}{r}{$\mathit{6.5}$} &  & 
		\multicolumn{1}{r}{$46.5$} &  \multicolumn{1}{r}{($12.8$)} & \multicolumn{1}{r}{$\mathit{23.2}$}   \\

		& 30 &  4 & 0.5 & 0.3 & & \multicolumn{1}{r}{$73.9$} & \multicolumn{1}{r}{($14.0$)} &
		\multicolumn{1}{r}{$\mathit{28.9}$} & & 
		\multicolumn{1}{r}{$72.2$} &  \multicolumn{1}{r}{($11.8$)} & \multicolumn{1}{r}{$\mathit{13.2}$}   \\

		& 30 &  4 & 0.5 & 0.8 & & \multicolumn{1}{r}{$48.6$} & \multicolumn{1}{r}{($13.6$)} &
		\multicolumn{1}{r}{$\mathit{17.3}$} &  &
		\multicolumn{1}{r}{$22.6$} &  \multicolumn{1}{r}{($4.4$)} & \multicolumn{1}{r}{$\mathit{9.3}$}   \\

		& 30 &  4 & 0.9 & 0.3 & & \multicolumn{1}{r}{$27.3$} & \multicolumn{1}{r}{($0.3$)} &
		\multicolumn{1}{r}{$\mathit{4.6}$} &  &
		\multicolumn{1}{r}{$87.5$} &  \multicolumn{1}{r}{($-0.7$)} & \multicolumn{1}{r}{$\mathit{2.4}$}   \\

		& 30 &  4 & 0.9 & 0.8 & & \multicolumn{1}{r}{$-63.3$} & \multicolumn{1}{r}{($6.2$)} &
		\multicolumn{1}{r}{$\mathit{4.3}$} &  &
		\multicolumn{1}{r}{$47.9$} &  \multicolumn{1}{r}{($12.8$)} & \multicolumn{1}{r}{$\mathit{25.7}$}   \\ 
		
		& & &  &  &  &  &  &   &  & & &  \\
		
		\multicolumn{8}{l}{\underline{Time-series heteroskedastic errors:}} & & & &   \\
		
		& & &  &  &  &  &  &   &  & & &  \\
		
		& 10 &  1 & 0.5 & 0.3 & & \multicolumn{1}{r}{$61.3$} & \multicolumn{1}{r}{($10.5$)} &
		\multicolumn{1}{r}{$\mathit{12.4}$} & & 
		\multicolumn{1}{r}{$80.0$} &  \multicolumn{1}{r}{($10.4$)} & \multicolumn{1}{r}{$\mathit{10.6}$}   \\

		& 10 &  1 & 0.5 & 0.8 & & \multicolumn{1}{r}{$56.7$} & \multicolumn{1}{r}{($12.6$)} &
		\multicolumn{1}{r}{$\mathit{13.5}$} & & 
		\multicolumn{1}{r}{$-1.1$} &  \multicolumn{1}{r}{($11.3$)} & \multicolumn{1}{r}{$\mathit{11.3}$}   \\

		& 10 &  1 & 0.9 & 0.3 & & \multicolumn{1}{r}{$63.9$} & \multicolumn{1}{r}{($25.0$)} &
		\multicolumn{1}{r}{$\mathit{31.9}$} &  &
		\multicolumn{1}{r}{$66.3$} &  \multicolumn{1}{r}{($19.3$)} & \multicolumn{1}{r}{$\mathit{23.4}$}   \\

		& 10 &  1 & 0.9 & 0.8 & & \multicolumn{1}{r}{$48.0$} & \multicolumn{1}{r}{($17.2$)} &
		\multicolumn{1}{r}{$\mathit{18.0}$} &  & 
		\multicolumn{1}{r}{$15.4$} &  \multicolumn{1}{r}{($14.0$)} & \multicolumn{1}{r}{$\mathit{14.0}$}   \\

		& 10 &  4 & 0.5 & 0.3 & & \multicolumn{1}{r}{$80.4$} & \multicolumn{1}{r}{($27.1$)} &
		\multicolumn{1}{r}{$\mathit{31.8}$} & & 
		\multicolumn{1}{r}{$95.5$} &  \multicolumn{1}{r}{($19.5$)} & \multicolumn{1}{r}{$\mathit{21.2}$}   \\

		& 10 &  4 & 0.5 & 0.8 & & \multicolumn{1}{r}{$77.4$} & \multicolumn{1}{r}{($24.7$)} &
		\multicolumn{1}{r}{$\mathit{26.6}$} &  &
		\multicolumn{1}{r}{$-275.3$} &  \multicolumn{1}{r}{($16.1$)} & \multicolumn{1}{r}{$\mathit{15.7}$}   \\

		& 10 &  4 & 0.9 & 0.3 & & \multicolumn{1}{r}{$82.8$} & \multicolumn{1}{r}{($47.4$)} &
		\multicolumn{1}{r}{$\mathit{57.7}$} &  &
		\multicolumn{1}{r}{$85.7$} &  \multicolumn{1}{r}{($39.5$)} & \multicolumn{1}{r}{$\mathit{49.2}$}   \\

		& 10 &  4 & 0.9 & 0.8 & & \multicolumn{1}{r}{$79.0$} & \multicolumn{1}{r}{($39.2$)} &
		\multicolumn{1}{r}{$\mathit{41.4}$} &  &
		\multicolumn{1}{r}{$93.9$} &  \multicolumn{1}{r}{($23.2$)} & \multicolumn{1}{r}{$\mathit{23.6}$}   \\

		& &  &  &  &  &  &  &   &  & & &  \\
		
		& 30 &  1 & 0.5 & 0.3 & & \multicolumn{1}{r}{$79.5$} & \multicolumn{1}{r}{($12.9$)} &
		\multicolumn{1}{r}{$\mathit{17.4}$} & & 
		\multicolumn{1}{r}{$70.6$} &  \multicolumn{1}{r}{($17.3$)} & \multicolumn{1}{r}{$\mathit{17.4}$}   \\

		& 30 &  1 & 0.5 & 0.8 & & \multicolumn{1}{r}{$79.3$} & \multicolumn{1}{r}{($15.3$)} &
		\multicolumn{1}{r}{$\mathit{16.6}$} & & 
		\multicolumn{1}{r}{$80.4$} &  \multicolumn{1}{r}{($22.6$)} & \multicolumn{1}{r}{$\mathit{23.3}$}   \\

		& 30 &  1 & 0.9 & 0.3 & & \multicolumn{1}{r}{$73.5$} & \multicolumn{1}{r}{($22.0$)} &
		\multicolumn{1}{r}{$\mathit{30.8}$} &  &
		\multicolumn{1}{r}{$75.7$} &  \multicolumn{1}{r}{($19.4$)} & \multicolumn{1}{r}{$\mathit{20.5}$}   \\

		& 30 &  1 & 0.9 & 0.8 & & \multicolumn{1}{r}{$51.4$} & \multicolumn{1}{r}{($16.8$)} &
		\multicolumn{1}{r}{$\mathit{17.4}$} &  & 
		\multicolumn{1}{r}{$85.2$} &  \multicolumn{1}{r}{($16.5$)} & \multicolumn{1}{r}{$\mathit{18.0}$}   \\

		& 30 &  4 & 0.5 & 0.3 & & \multicolumn{1}{r}{$90.6$} & \multicolumn{1}{r}{($25.8$)} &
		\multicolumn{1}{r}{$\mathit{35.3}$} & & 
		\multicolumn{1}{r}{$98.5$} &  \multicolumn{1}{r}{($22.4$)} & \multicolumn{1}{r}{$\mathit{24.3}$}   \\

		& 30 &  4 & 0.5 & 0.8 & & \multicolumn{1}{r}{$91.0$} & \multicolumn{1}{r}{($25.7$)} &
		\multicolumn{1}{r}{$\mathit{29.3}$} &  &
		\multicolumn{1}{r}{$24.0$} &  \multicolumn{1}{r}{($26.7$)} & \multicolumn{1}{r}{$\mathit{26.7}$}   \\

		& 30 &  4 & 0.9 & 0.3 & & \multicolumn{1}{r}{$94.0$} & \multicolumn{1}{r}{($53.8$)} &
		\multicolumn{1}{r}{$\mathit{69.9}$} &  &
		\multicolumn{1}{r}{$97.3$} &  \multicolumn{1}{r}{($40.9$)} & \multicolumn{1}{r}{$\mathit{54.7}$}   \\

		& 30 &  4 & 0.9 & 0.8 & & \multicolumn{1}{r}{$86.8$} & \multicolumn{1}{r}{($36.7$)} &
		\multicolumn{1}{r}{$\mathit{40.3}$} &  &
		\multicolumn{1}{r}{$-21.3$} &  \multicolumn{1}{r}{($23.0$)} & \multicolumn{1}{r}{$\mathit{23.0}$}   \\

		&	&  &  &  &  &  &  &   &  &   &   & \\ \hline
	\end{tabular}
	\label{fod_vs_fd} 
\end{table}

\begin{table}[H]
	\caption{Bias, standard deviation, and root mean squared error estimates for FD-SYS  ($N=$ $200$).}
	\centering
	
	\begin{tabular}{lcccccccccc}
		\hline
		&  & &  & &  \multicolumn{3}{c}{$\delta $ estimator} &    \multicolumn{3}{c}{$\alpha$ estimator}    \\ 
		&  $ T $ & $\sigma _{\eta }$ & $\delta$ & $\rho$ & bias & (sd) &
		\textit{rmse} &   bias & (sd) & \textit{rmse}  \\ \hline
		&  &  &  &  &  &  &   &  &  & \\ 
		
		\multicolumn{6}{l}{\underline{Conditionally heteroskedastic errors:}} & & & &  \\
		
		&  &  &  &  &  &  &   &  &  & \\

		& $10$ &  1 & 0.5 & 0.3 &  \multicolumn{1}{r}{$-0.0053$} & \multicolumn{1}{r}{($0.0338$)} &
		\multicolumn{1}{r}{$\mathit{0.0342}$} &  
		\multicolumn{1}{r}{$0.0026$} &  \multicolumn{1}{r}{($0.0409$)} & \multicolumn{1}{r}{$\mathit{0.0409}$}   \\

		& $10$ & 1 & 0.5 & 0.8 &  \multicolumn{1}{r}{$-0.0068$} & \multicolumn{1}{r}{($0.0316$)} &
		\multicolumn{1}{r}{$\mathit{0.0324}$} &  
		\multicolumn{1}{r}{$0.0286$} &  \multicolumn{1}{r}{($0.0500$)} & \multicolumn{1}{r}{$\mathit{0.0575}$}   \\

		& $10$ & 1 & 0.9 & 0.3 &  \multicolumn{1}{r}{$-0.0531$} & \multicolumn{1}{r}{($0.0849$)} &
		\multicolumn{1}{r}{$\mathit{0.1001}$} &  
		\multicolumn{1}{r}{$0.0072$} &  \multicolumn{1}{r}{($0.0674$)} & \multicolumn{1}{r}{$\mathit{0.0678}$}   \\

		& $10$ & 1 & 0.9 & 0.8 &  \multicolumn{1}{r}{$-0.0459$} & \multicolumn{1}{r}{($0.1366$)} &
		\multicolumn{1}{r}{$\mathit{0.1441}$} &  
		\multicolumn{1}{r}{$0.1186$} &  \multicolumn{1}{r}{($0.2431$)} & \multicolumn{1}{r}{$\mathit{0.2705}$}   \\ 
		
		&  &  &  &  &  &  &   &  &   &    \\

		& $10$ & 4 & 0.5 & 0.3 &  \multicolumn{1}{r}{$-0.0279$} & \multicolumn{1}{r}{($0.0504$)} &
		\multicolumn{1}{r}{$\mathit{0.0576}$} &  
		\multicolumn{1}{r}{$-0.0060$} &  \multicolumn{1}{r}{($0.0483$)} & \multicolumn{1}{r}{$\mathit{0.0486}$}   \\

		& $10$ & 4 & 0.5 & 0.8 &  \multicolumn{1}{r}{$-0.0138$} & \multicolumn{1}{r}{($0.0452$)} &
		\multicolumn{1}{r}{$\mathit{0.0473}$} &  
		\multicolumn{1}{r}{$0.0350$} &  \multicolumn{1}{r}{($0.0606$)} & \multicolumn{1}{r}{$\mathit{0.0700}$}   \\

		& $10$ & 4 & 0.9 & 0.3 &  \multicolumn{1}{r}{$-0.1221$} & \multicolumn{1}{r}{($0.1714$)} &
		\multicolumn{1}{r}{$\mathit{0.2105}$} &  
		\multicolumn{1}{r}{$-0.0206$} &  \multicolumn{1}{r}{($0.2479$)} & \multicolumn{1}{r}{$\mathit{0.2487}$}   \\ 
		
		& $10$ & 4 & 0.9 & 0.8 &  \multicolumn{1}{r}{$-0.0501$} & \multicolumn{1}{r}{($0.1544$)} &
		\multicolumn{1}{r}{$\mathit{0.1623}$} &  
		\multicolumn{1}{r}{$0.1520$} &  \multicolumn{1}{r}{($0.2705$)} & \multicolumn{1}{r}{$\mathit{0.3103}$}   \\ 
		
		&  &  &  &  &  &  &   &  &  & \\ 
		
		\multicolumn{6}{l}{\underline{Time-series heteroskedastic errors:}} & & & &  \\	
		
		&  &  &  &  &  &  &   &  &  & \\

		& $10$ & 1 & 0.5 & 0.3 &  \multicolumn{1}{r}{$-0.0024$} & \multicolumn{1}{r}{($0.0216$)} &
		\multicolumn{1}{r}{$\mathit{0.0217}$} &  
		\multicolumn{1}{r}{$0.0004$} &  \multicolumn{1}{r}{($0.0223$)} & \multicolumn{1}{r}{$\mathit{0.0223}$}   \\

		& $10$ & 1 & 0.5 & 0.8 &  \multicolumn{1}{r}{$-0.0010$} & \multicolumn{1}{r}{($0.0166$)} &
		\multicolumn{1}{r}{$\mathit{0.0167}$} &  
		\multicolumn{1}{r}{$0.0015$} &  \multicolumn{1}{r}{($0.0197$)} & \multicolumn{1}{r}{$\mathit{0.0198}$}   \\

		& $10$ & 1 & 0.9 & 0.3 &  \multicolumn{1}{r}{$-0.0147$} & \multicolumn{1}{r}{($0.0326$)} &
		\multicolumn{1}{r}{$\mathit{0.0358}$} &  
		\multicolumn{1}{r}{$-0.0068$} &  \multicolumn{1}{r}{($0.0256$)} & \multicolumn{1}{r}{$\mathit{0.0264}$}   \\ 
		
		& $10$ & 1 & 0.9 & 0.8 &  \multicolumn{1}{r}{$-0.0011$} & \multicolumn{1}{r}{($0.0120$)} &
		\multicolumn{1}{r}{$\mathit{0.0120}$} &  
		\multicolumn{1}{r}{$0.0002$} &  \multicolumn{1}{r}{($0.0134$)} & \multicolumn{1}{r}{$\mathit{0.0134}$}   \\ 
		
		&  &  &  &  &  &  & &  &  & \\ 
		
		& $10$ & 4 & 0.5 & 0.3 &  \multicolumn{1}{r}{$-0.0122$} & \multicolumn{1}{r}{($0.0338$)} &
		\multicolumn{1}{r}{$\mathit{0.0359}$} &  
		\multicolumn{1}{r}{$-0.0042$} &  \multicolumn{1}{r}{($0.0285$)} & \multicolumn{1}{r}{$\mathit{0.0288}$}   \\ 
		
		& $10$ & 4 & 0.5 & 0.8 &  \multicolumn{1}{r}{$-0.0036$} & \multicolumn{1}{r}{($0.0223$)} &
		\multicolumn{1}{r}{$\mathit{0.0226}$} &  
		\multicolumn{1}{r}{$0.0016$} &  \multicolumn{1}{r}{($0.0230$)} & \multicolumn{1}{r}{$\mathit{0.0230}$}   \\ 
		
		& $10$ & 4 & 0.9 & 0.3 &  \multicolumn{1}{r}{$-0.0657$} & \multicolumn{1}{r}{($0.0704$)} &
		\multicolumn{1}{r}{$\mathit{0.0963}$} &  
		\multicolumn{1}{r}{$-0.0314$} &  \multicolumn{1}{r}{($0.0431$)} & \multicolumn{1}{r}{$\mathit{0.0533}$}   \\ 
		
		& $10$ & 4 & 0.9 & 0.8 &  \multicolumn{1}{r}{$-0.0042$} & \multicolumn{1}{r}{($0.0176$)} &
		\multicolumn{1}{r}{$\mathit{0.0180}$} &  
		\multicolumn{1}{r}{$-0.0010$} &  \multicolumn{1}{r}{($0.0155$)} & \multicolumn{1}{r}{$\mathit{0.0155}$}   \\ 
		
		&  &  &  &  &  &  &   &   &  &    \\ \hline
	\end{tabular} 
	\label{fdsys_bias_sd_rmse}
\end{table}

\begin{table}[H]
	\caption{Percent reduction in absolute bias, standard deviation, and root mean squared error for FOD-SYS relative to FD-SYS (100(FD-SYS $-$ FOD-SYS)/FD-SYS). }
	\centering
	\begin{tabular}{lcccccccccccc}
		\hline
		& & & &  & &  \multicolumn{3}{c}{$\delta $ estimator} & &   \multicolumn{3}{c}{$\alpha$ estimator}    \\ 
		& $T$ &  $\sigma _{\eta }$ & $\delta$ & $\rho$ & & bias & (sd) &
		\textit{rmse} & &  bias & sd & \textit{rmse}  \\ \hline
		& &  &  &  &  &  &  &   &  & & &  \\
		
		\multicolumn{8}{l}{\underline{Conditionally heteroskedastic errors:}} & & & &   \\
		
		& & &  &  &  &  &  &   &  & & &  \\
		
		& 10 &  1 & 0.5 & 0.3 & & \multicolumn{1}{r}{$43.1$} & \multicolumn{1}{r}{($11.9$)} &
		\multicolumn{1}{r}{$\mathit{12.5}$} & & 
		\multicolumn{1}{r}{$-52.8$} &  \multicolumn{1}{r}{($6.0$)} & \multicolumn{1}{r}{$\mathit{5.7}$}   \\

		& 10 &  1 & 0.5 & 0.8 & & \multicolumn{1}{r}{$27.3$} & \multicolumn{1}{r}{($14.1$)} &
		\multicolumn{1}{r}{$\mathit{14.7}$} & & 
		\multicolumn{1}{r}{$1.9$} &  \multicolumn{1}{r}{($4.3$)} & \multicolumn{1}{r}{$\mathit{3.7}$}   \\

		& 10 &  1 & 0.9 & 0.3 & & \multicolumn{1}{r}{$24.6$} & \multicolumn{1}{r}{($15.9$)} &
		\multicolumn{1}{r}{$\mathit{18.2}$} &  &
		\multicolumn{1}{r}{$-113.4$} &  \multicolumn{1}{r}{($15.1$)} & \multicolumn{1}{r}{$\mathit{12.6}$}   \\

		& 10 &  1 & 0.9 & 0.8 & & \multicolumn{1}{r}{$6.8$} & \multicolumn{1}{r}{($11.5$)} &
		\multicolumn{1}{r}{$\mathit{11.0}$} &  & 
		\multicolumn{1}{r}{$17.4$} &  \multicolumn{1}{r}{($8.1$)} & \multicolumn{1}{r}{$\mathit{9.8}$}   \\

		& & &  &  &  &  &  &   &  & & & \\ 
		
		& 10 &  4 & 0.5 & 0.3 & & \multicolumn{1}{r}{$76.7$} & \multicolumn{1}{r}{($23.1$)} &
		\multicolumn{1}{r}{$\mathit{31.8}$} & & 
		\multicolumn{1}{r}{$15.2$} &  \multicolumn{1}{r}{($9.8$)} & \multicolumn{1}{r}{$\mathit{9.8}$}   \\

		& 10 &  4 & 0.5 & 0.8 & & \multicolumn{1}{r}{$56.5$} & \multicolumn{1}{r}{($17.8$)} &
		\multicolumn{1}{r}{$\mathit{20.4}$} &  &
		\multicolumn{1}{r}{$-2.1$} &  \multicolumn{1}{r}{($3.3$)} & \multicolumn{1}{r}{$\mathit{1.9}$}   \\

		& 10 &  4 & 0.9 & 0.3 & & \multicolumn{1}{r}{$48.6$} & \multicolumn{1}{r}{($19.7$)} &
		\multicolumn{1}{r}{$\mathit{28.1}$} &  &
		\multicolumn{1}{r}{$-42.2$} &  \multicolumn{1}{r}{($8.1$)} & \multicolumn{1}{r}{$\mathit{7.7}$}   \\

		& 10 &  4 & 0.9 & 0.8 & & \multicolumn{1}{r}{$11.5$} & \multicolumn{1}{r}{($13.7$)} &
		\multicolumn{1}{r}{$\mathit{13.5}$} &  &
		\multicolumn{1}{r}{$18.8$} &  \multicolumn{1}{r}{($9.5$)} & \multicolumn{1}{r}{$\mathit{11.6}$}   \\ 
		
		& &  &  &  &  &  &  &   &  & & &  \\
		
		\multicolumn{8}{l}{\underline{Time-series heteroskedastic errors:}} & & & &   \\
		
		& & &  &  &  &  &  &   &  & & &  \\
		
		& 10 &  1 & 0.5 & 0.3 & & \multicolumn{1}{r}{$71.8$} & \multicolumn{1}{r}{($9.7$)} &
		\multicolumn{1}{r}{$\mathit{10.2}$} & & 
		\multicolumn{1}{r}{$-154.3$} &  \multicolumn{1}{r}{($7.9$)} & \multicolumn{1}{r}{$\mathit{7.8}$}   \\

		& 10 &  1 & 0.5 & 0.8 & & \multicolumn{1}{r}{$64.8$} & \multicolumn{1}{r}{($9.2$)} &
		\multicolumn{1}{r}{$\mathit{9.4}$} & & 
		\multicolumn{1}{r}{$2.4$} &  \multicolumn{1}{r}{($8.9$)} & \multicolumn{1}{r}{$\mathit{8.8}$}   \\

		& 10 &  1 & 0.9 & 0.3 & & \multicolumn{1}{r}{$69.1$} & \multicolumn{1}{r}{($24.4$)} &
		\multicolumn{1}{r}{$\mathit{29.9}$} &  &
		\multicolumn{1}{r}{$73.0$} &  \multicolumn{1}{r}{($14.3$)} & \multicolumn{1}{r}{$\mathit{16.9}$}   \\

		& 10 &  1 & 0.9 & 0.8 & & \multicolumn{1}{r}{$43.8$} & \multicolumn{1}{r}{($11.1$)} &
		\multicolumn{1}{r}{$\mathit{11.3}$} &  & 
		\multicolumn{1}{r}{$-9.0$} &  \multicolumn{1}{r}{($7.8$)} & \multicolumn{1}{r}{$\mathit{7.8}$}   \\

		& & &  &  &  &  &  &   &  & & & \\ 
		
		& 10 &  4 & 0.5 & 0.3 & & \multicolumn{1}{r}{$80.9$} & \multicolumn{1}{r}{($25.8$)} &
		\multicolumn{1}{r}{$\mathit{29.9}$} & & 
		\multicolumn{1}{r}{$85.9$} &  \multicolumn{1}{r}{($16.3$)} & \multicolumn{1}{r}{$\mathit{17.2}$}   \\

		& 10 &  4 & 0.5 & 0.8 & & \multicolumn{1}{r}{$85.5$} & \multicolumn{1}{r}{($20.6$)} &
		\multicolumn{1}{r}{$\mathit{21.6}$} &  &
		\multicolumn{1}{r}{$-48.4$} &  \multicolumn{1}{r}{($12.2$)} & \multicolumn{1}{r}{$\mathit{11.8}$}   \\

		& 10 &  4 & 0.9 & 0.3 & & \multicolumn{1}{r}{$81.1$} & \multicolumn{1}{r}{($46.5$)} &
		\multicolumn{1}{r}{$\mathit{58.8}$} &  &
		\multicolumn{1}{r}{$85.8$} &  \multicolumn{1}{r}{($36.3$)} & \multicolumn{1}{r}{$\mathit{47.8}$}   \\

		& 10 &  4 & 0.9 & 0.8 & & \multicolumn{1}{r}{$78.0$} & \multicolumn{1}{r}{($27.5$)} &
		\multicolumn{1}{r}{$\mathit{29.3}$} &  &
		\multicolumn{1}{r}{$97.0$} &  \multicolumn{1}{r}{($11.5$)} & \multicolumn{1}{r}{$\mathit{11.7}$}   \\

		& & &  &  &  &  &  &   &  &   &   & \\ \hline
	\end{tabular} 
	\label{fodsys_vs_fdsys}
\end{table}

\section{Summary \label{summary}}

This paper showed that the necessary and sufficient instruments condition provided in \cite{Phillips2019a} applies to two-step GMM estimation based on heteroskedasticity-robust weighting matrices.  If the condition is satisfied, a two-step GMM estimator, based on an optimal weighting matrix, can be calculated using another transformation and the optimal weighting matrix corresponding to the alternative transformation.   The paper also showed when the system GMM estimator studied by \cite{Bover1995} and \cite{Blundell1998} can be computed using forward orthogonal deviations rather than first differencing.

Because the instruments condition is not just sufficient but also necessary, it tells us when GMM estimators are not invariant to transformation. One situation for which invariance to transformation is not possible is when only recent lags of predetermined variables are used as instrumental variables.  

Monte Carlo experiments were used to examine two important cases: two-step FD-GMM estimation versus two-step FOD-GMM estimation and FD-SYS estimation versus FOD-SYS estimation. When these GMM estimators exploited only recent lags of predetermined variables as instruments, the estimators based on forward orthogonal deviations were generally less biased and almost always more efficient than their counterparts based on first differencing. 

\section{Proofs \label{proofs}}

The proof of Theorem \ref{thm1} relies on a corollary to Theorem 1 in \cite{Phillips2019a}. That corollary is stated as Lemma \ref{lemma1} here.

\begin{lemma} \label{lemma1}
	Let $\boldsymbol{U}$ be the upper-triangular Cholesky factorization of $(\boldsymbol{K}\boldsymbol{K}^{\prime})^{-1}$, and let $\boldsymbol{Z}_i$ be defined as in (\ref{Z_i}).  Then there is a nonsingular matrix $\boldsymbol{C}$ satisfying $\boldsymbol{C}\boldsymbol{Z}_i^{\prime} = \boldsymbol{Z}_i^{\prime}\boldsymbol{U}$ if, and only if, every entry in $\boldsymbol{z}_{is}$ is a linear combination of the entries in $\boldsymbol{z}_{it}$ ($s =1,\ldots, t$, $t=1,\ldots,R$).
\end{lemma}

\noindent \textit{Proof of Lemma \ref{lemma1}}: The conclusion of the lemma follows from Theorem 1 in \cite{Phillips2019a}.  To apply that theorem, let $\boldsymbol{\Phi} = \boldsymbol{K}\boldsymbol{\Omega}\boldsymbol{K}^{\prime}$, where $\boldsymbol{\Omega}$ is a positive definite matrix. Theorem 1 in \cite{Phillips2019a} says that for the upper-triangular Cholesky factorization of $\boldsymbol{\Phi}^{-1}$, say $\boldsymbol{U}^{\ast}$, there is a nonsingular matrix $\boldsymbol{C}$ satisfying $\boldsymbol{C}\boldsymbol{Z}_i^{\prime} = \boldsymbol{Z}_i^{\prime}\boldsymbol{U}^{\ast}$ if, and only if, every entry in $\boldsymbol{z}_{is}$ is a linear combination of the entries in $\boldsymbol{z}_{it}$ ($s = 1,\ldots, t$, $t=1,\ldots,R$).  Set $\boldsymbol{\Omega}=\boldsymbol{I}$. Then $ \boldsymbol{U}^{\ast} = \boldsymbol{U} $, and the conclusion of Lemma \ref{lemma1} follows. \\

\subsection{Proof of Theorem \ref{thm1}}

Under the conditions of the theorem, we have by Lemma \ref{lemma1} that there is a nonsingular matrix $\boldsymbol{C}$ such that $\boldsymbol{C}\boldsymbol{Z}_i^{\prime}=\boldsymbol{Z}_i^{\prime}\boldsymbol{U}$ if, and only if, every entry in $\boldsymbol{z}_{is}$ is a linear combination of entries in $\boldsymbol{z}_{it}$ $(s = 1, \ldots, t$, $%
t=1,\ldots ,R)$. This fact and $\boldsymbol{F}=\boldsymbol{U}\boldsymbol{K}$ imply 
\begin{eqnarray}
\sum_i\boldsymbol{\ddot{X}}%
_{i}^{ \prime }\boldsymbol{Z}_{i}\left( \sum_i\boldsymbol{Z}_{i}^{\prime }\boldsymbol{\ddot{e}}_{i}\boldsymbol{\ddot{e}}_{i}^{\prime}\boldsymbol{Z}_i\right) ^{-1} 
 \sum_i 
\boldsymbol{Z}%
_{i}^{\prime }\boldsymbol{\ddot{X}}_{i}  
 & =&  \sum_i\boldsymbol{\tilde{X}}%
_{i}^{ \prime }\boldsymbol{U}^{\prime}\boldsymbol{Z}_{i} \left( \sum_i\boldsymbol{Z}_{i}^{\prime }\boldsymbol{U}\boldsymbol{\tilde{e}}_{i}\boldsymbol{\tilde{e}}_{i}^{\prime}\boldsymbol{U}^{\prime}\boldsymbol{Z}_i\right) ^{-1}\sum_i\boldsymbol{Z}%
_{i}^{\prime }\boldsymbol{U}\boldsymbol{\tilde{X}}_{i} \notag \\
& = & \sum_i\boldsymbol{\tilde{X}}%
_{i}^{ \prime }\boldsymbol{Z}_{i}\boldsymbol{C}^{\prime} \left( \sum_i\boldsymbol{C}\boldsymbol{Z}_{i}^{\prime }\boldsymbol{\tilde{e}}_{i}\boldsymbol{\tilde{e}}_{i}^{\prime}\boldsymbol{Z}_i\boldsymbol{C}^{\prime}\right) ^{-1}\sum_i\boldsymbol{C}\boldsymbol{Z}%
_{i}^{\prime }\boldsymbol{\tilde{X}}_{i} \notag \\
& = & \sum_i\boldsymbol{\tilde{X}}%
_{i}^{ \prime }\boldsymbol{Z}_{i} \left( \sum_i\boldsymbol{Z}_{i}^{\prime }\boldsymbol{\tilde{e}}_{i}\boldsymbol{\tilde{e}}_{i}^{\prime}\boldsymbol{Z}_i\right) ^{-1}\sum_i\boldsymbol{Z}%
_{i}^{\prime }\boldsymbol{\tilde{X}}_{i} \label{der1}
\end{eqnarray}
if, and only if, every entry in $\boldsymbol{z}_{is}$ is a linear combination of entries in $\boldsymbol{z}_{it}$ $(s = 1, \ldots, t$, $%
t=1,\ldots ,R)$. By similar reasoning, we get
\begin{equation}
\sum_i\boldsymbol{\ddot{X}}%
_{i}^{ \prime }\boldsymbol{Z}_{i}\left( \sum_i\boldsymbol{Z}_{i}^{\prime }\boldsymbol{\ddot{e}}_{i}\boldsymbol{\ddot{e}}_{i}^{\prime}\boldsymbol{Z}_i\right) ^{-1} 
 \sum_i 
\boldsymbol{Z}%
_{i}^{\prime }\boldsymbol{\ddot{y}}_{i}  
 =  \sum_i\boldsymbol{\tilde{X}}%
_{i}^{ \prime }\boldsymbol{Z}_{i} \left( \sum_{i=1}^{n}\boldsymbol{Z}_{i}^{\prime }\boldsymbol{\tilde{e}}_{i}\boldsymbol{\tilde{e}}_{i}^{\prime}\boldsymbol{Z}_i\right) ^{-1}\sum_{i=1}^{n}\boldsymbol{Z}%
_{i}^{\prime }\boldsymbol{\tilde{y}}_{i}, \label{der2}
\end{equation}
if, and only if, every entry in $\boldsymbol{z}_{is}$ is a linear combination of entries in $\boldsymbol{z}_{it}$ $(s = 1, \ldots, t$, $%
t=1,\ldots ,R)$.

\subsection{Proof of Theorem \ref{thm2}}

By Lemma \ref{lemma1}, there is a nonsingular matrix $\boldsymbol{C}$ satisfying $\boldsymbol{C}\boldsymbol{Z}_{1i}^{\prime}=\boldsymbol{Z}_{1i}^{\prime}\boldsymbol{U}$ if, and only if, every entry in $\boldsymbol{z}_{is}$ is a linear combination of entries in $\boldsymbol{z}_{it}$ $(s = 1, \ldots, t$, $%
t=1,\ldots ,T-1)$. Let 
\begin{equation*}
\boldsymbol{C}^+=\left( 
\begin{array}{cc}
\boldsymbol{C} & \boldsymbol{0}    \\ 
\boldsymbol{0} & \boldsymbol{I}  \\ 
\end{array}%
\right) \label{C+}
\end{equation*}
and
\begin{equation*}
\boldsymbol{U}^+=\left( 
\begin{array}{cc}
\boldsymbol{U} & \boldsymbol{0}    \\ 
\boldsymbol{0} & \boldsymbol{I}  \\ 
\end{array}%
\right). \label{U+}
\end{equation*}
Then $\boldsymbol{C}^+$ is nonsingular and $\boldsymbol{C}^+\boldsymbol{Z}_i^{+ \prime}=\boldsymbol{Z}_i^{+ \prime}\boldsymbol{U}^+$ if, and only if, every entry in $\boldsymbol{z}_{is}$ is a linear combination of entries in $\boldsymbol{z}_{it}$ $(s = 1,\ldots, t$, $%
t=1,\ldots ,T-1)$. This fact and $\boldsymbol{F}^+=\boldsymbol{U}^+\boldsymbol{K}^+$ gives
\begin{equation*}
\sum_i\boldsymbol{\ddot{X}}%
_{i}^{+ \prime }\boldsymbol{Z}_{i}^+\left( \sum_i\boldsymbol{Z}_{i}^{+ \prime }\boldsymbol{\ddot{e}}_{i}^+\boldsymbol{\ddot{e}}_{i}^{+ \prime}\boldsymbol{Z}_i^+\right) ^{-1} 
 \sum_i 
\boldsymbol{Z}%
_{i}^{+ \prime }\boldsymbol{\ddot{X}}_{i}^+  
 =  \sum_i\boldsymbol{\tilde{X}}%
_{i}^{+ \prime }\boldsymbol{Z}_{i}^+ \left( \sum_i\boldsymbol{Z}_{i}^{+ \prime }\boldsymbol{\tilde{e}}_{i}^+\boldsymbol{\tilde{e}}_{i}^{+ \prime}\boldsymbol{Z}_i^+\right) ^{-1}\sum_i\boldsymbol{Z}%
_{i}^{+ \prime }\boldsymbol{\tilde{X}}_{i}^+ 
\end{equation*}
and
\begin{equation*}
\sum_i\boldsymbol{\ddot{X}}%
_{i}^{+ \prime }\boldsymbol{Z}_{i}^+\left( \sum_i\boldsymbol{Z}_{i}^{+ \prime }\boldsymbol{\ddot{e}}_{i}^+\boldsymbol{\ddot{e}}_{i}^{+ \prime}\boldsymbol{Z}_i^+\right) ^{-1} 
 \sum_i 
\boldsymbol{Z}%
_{i}^{+ \prime }\boldsymbol{\ddot{y}}_{i}^+  
 =  \sum_i\boldsymbol{\tilde{X}}%
_{i}^{+ \prime }\boldsymbol{Z}_{i}^+ \left( \sum_i\boldsymbol{Z}_{i}^{+ \prime }\boldsymbol{\tilde{e}}_{i}^+\boldsymbol{\tilde{e}}_{i}^{+ \prime}\boldsymbol{Z}_i^+\right) ^{-1}\sum_i\boldsymbol{Z}%
_{i}^{+ \prime }\boldsymbol{\tilde{y}}_{i}^+ 
\end{equation*}
by derivations similar to those establishing Eq.s (\ref{der1}) and (\ref{der2}).

\end{document}